\documentclass[12pt]{article}
\usepackage[dvips]{graphicx}

\usepackage{epsfig,amsmath,amssymb,verbatim,mathrsfs}

\def\beq{\begin{eqnarray}}
\def\eeq{\end{eqnarray}}
\def\bea{\begin{eqnarray}}
\def\eea{\end{eqnarray}}
\def\blfootnote{\xdef\@thefnmark{}\@footnotetext}

\newcommand{\yp}{y_\phi}

\newcommand{\be}{\begin{equation}}
\newcommand{\ee}{\end{equation}}

\begin{document}
\begin{titlepage}
\noindent

\setlength{\baselineskip}{0.2in}
\flushright{NIKHEF/2011-020\\July 2011}


\begin{center}
  \begin{Large}
    \begin{bf}
  Gravity on a Little Warped Space  
\end{bf}
  \end{Large}
\end{center}
\vspace{0.2cm}

\begin{center}
\begin{large}
{
  Damien~P.~George$^{(a),}$\footnote{Email: dpgeorge@nikhef.nl} and Kristian
  L.~McDonald$^{(b),}$\footnote{Email: kristian.mcdonald@mpi-hd.mpg.de}}\\
\end{large}
\vspace{1cm}
  \begin{it}
$(a)$ Nikhef Theory Group, Science Park 105,\\
1098 XG Amsterdam, The Netherlands\\\vspace{0.5cm}
$(b)$ Max-Planck-Institut f\"ur Kernphysik,\\
 Postfach 10 39 80, 69029 Heidelberg, Germany.\vspace{0.5cm}\\
\end{it}
\vspace{1cm}

\end{center}


\begin{abstract}
We investigate the consistent inclusion of 4D Einstein gravity on a
truncated slice of $AdS_5$ whose bulk-gravity and UV
scales are much less than the 4D Planck scale, $M_*\ll
M_{Pl}$. Such ``Little Warped Spaces'' have found
phenomenological utility and can be motivated by string
realizations of the Randall-Sundrum framework. Using the
interval approach to brane-world gravity, we show that the inclusion
of a large UV-localized
Einstein-Hilbert term allows one to consistently incorporate 4D Einstein
gravity into the low-energy theory. We detail the spectrum of
Kaluza-Klein metric fluctuations and, in particular, examine the
coupling of the little radion to matter. Furthermore, we show that
Goldberger-Wise stabilization can be successfully implemented on such
spaces.  Our
results  demonstrate that realistic low-energy effective theories can be constructed on these spaces, and have
relevance for existing models in the literature.
\end{abstract}

\vspace{2cm}

\end{titlepage}

\setcounter{page}{1}
\setcounter{footnote}{0}


\vfill\eject


\section{Introduction}
The Randall-Sundrum (RS) framework provides a natural means by which to generate hierarchically separated, radiatively-stable mass scales~\cite{Randall:1999ee}. As such, it has received much attention in connection with the gauge hierarchy problem of the Standard Model (SM). In its standard incarnation, the ultraviolet (UV) scale of a warped extra dimension is (approximately) identified with the 4D Planck scale, while the infrared (IR) scale is associated with electroweak symmetry breaking (i.e.\ the weak scale). The hierarchy between these two scales is generated by spacetime warping, resulting in a mechanism for realizing the weak scale from Planck-scale sized input parameters.

Of course, any solution to the gauge hierarchy problem should explain why the weak scale is radiatively stable relative to some UV cutoff, $M_*\gtrsim$~TeV. However, this cutoff need not be the Planck scale and may instead be some intermediate scale at which  new physics arises in connection with, e.g., baryogenesis, flavor, etc. This is precisely the philosophy of the Little Randall-Sundrum (LRS) model~\cite{Davoudiasl:2008hx,Bauer:2008xb,Davoudiasl:2010fb}, which employs a Little Warped Space (LWS)~\cite{McDonald:2010jm, Duerr:2011ks} --- namely a truncated slice of $AdS_5$ whose bulk-gravity and UV scales are much less than the 4D Planck mass, $M_*\ll M_{Pl}$ --- to realize a candidate solution to the hierarchy problem. This framework provides an explanation for the stability of the weak scale but only relative to some intermediate scale $M_*$, beyond which a UV completion is expected. 

If the UV scale of a warped space is decoupled from the 4D Planck scale, an evident question arises: How does one include 4D Einstein gravity in such a low energy effective theory framework? A necessary ingredient for any realistic low energy theory is that it should reproduce 4D Einstein gravity at large distances, a point that must also hold  for models constructed on a LWS. Despite the fact that such theories are not intended to explain the full weak/Planck hierarchy, it should be possible to incorporate 4D gravity if the effective theory framework is to be valid. To date, this matter has not been addressed in the literature.\footnote{A previous work has embedded the LRS model in an extended spacetime~\cite{McDonald:2008ss}.} It is the goal of the present work to study the consistent inclusion of 4D gravity in models constructed on a LWS, and to detail the spectrum of physical metric fluctuations obtained once 4D gravity is included.

As suggested in Ref.~\cite{McDonald:2010jm}, the key ingredient that enables 4D Einstein gravity to be realized on a LWS is the inclusion of a large UV-localized Einstein-Hilbert term.\footnote{Equivalently, a UV-localized ``brane curvature'' or ``DGP'' term. We use these labels for the UV term  interchangeably throughout this work.} Large brane-localized  curvature terms were originally studied by Dvali, Gabadadze, and Porrati (DGP)~\cite{Dvali:2000hr} in the context of a flat extra dimension, and have a rich history in the literature. A number of works have studied the issues of stability and
consistency that arise in the presence of a 
DGP term, and some subtleties have been found~\cite{Nicolis:2004qq,Gorbunov:2005zk,Adams:2006sv,Charmousis:2006pn} concerning, e.g., the onset of strong
coupling among metric fluctuations. However, it is known that no such strong coupling problems occur when a large UV localized DGP term is introduced in warped models, at least for energies below the original UV cutoff of the theory~\cite{Luty:2003vm}.

There exists an additional formal reason to consider the effects of a
large UV localized Einstein-Hilbert term in warped models. As is well
known, strongly warped regions that admit an RS-like description can arise
naturally in string theory flux
compactifications~\cite{Verlinde:1999fy}. Interestingly, UV-localized curvature terms that are parametrically large relative
to the effective UV scale can easily arise in the string
constructs~\cite{Brummer:2005sh}. This suggests that the approach of
the LRS model~\cite{Davoudiasl:2008hx} and the
Mini-Seesaw~\cite{McDonald:2010jm, Duerr:2011ks} (which also employs a
LWS) may be well motivated
from a ``top-down'' perspective. However, the broader study of phenomenological models on a LWS requires
one to understand the gravitational sector of these spaces in detail, and to
demonstrate their stability. Our aim is to carry out these analyses
and thus demonstrate conclusively that viable low
energy models which include Einstein gravity can be realized.

Throughout this work we consider warped spaces with a low UV scale, $M_*\ll M_{Pl}$, but for the most part we will not specify the particular value of $M_*$. Previous works employing a LWS have taken the UV scale to be \emph{significantly} lighter than the Planck scale; in the case of the LRS model one has $M_*\sim10^3$~TeV, while in Refs.~\cite{McDonald:2010jm, Duerr:2011ks}  the UV scale is taken at the TeV scale (with the warping employed to generate a light hidden sector scale). Our analysis applies to these works, but applies more generally to models where the hierarchy between $M_*$ and $M_{Pl}$ is not so severe. For example, our results would be relevant for models with a UV scale of $M_*\sim10^{14}$~GeV in connection with, e.g., a Grand Unification scale. Such models still require a parametrically large UV curvature term to reproduce viable 4D gravity, so the relation $M_*\ll M_{Pl}$ still holds. Given that string realizations of RS models readily generate large UV curvature terms~\cite{Brummer:2005sh}, it would seem prudent to consider the UV scale of the RS framework as a free parameter that may or may not be directly connected to the 4D Planck scale, and to study the phenomenological consequences of its variation. Our results remain relevant in this context.

The layout of this work is as follows. In Sec.~\ref{sec:setup} we
describe the setup for our analysis and derive the spectrum of
spin-two Kaluza-Klein (KK) modes in the presence of the large UV
Einstein-Hilbert term. We focus on the radion in
Secs.~\ref{sec:_massless_radion}
and~\ref{sec:radion-matter-GI}, where we, respectively, obtain the
precise form of the physical gravi-scalar fluctuation in the presence
of the UV term, and consider the
coupling of the ``little radion'' to matter. In
Sec.~\ref{sec:little_GW} we show that the length of a LWS (with
large UV brane term) is readily stabilized by implementing the
Goldberger-Wise mechanism~\cite{Goldberger:1999uk}. We obtain the
radion mass and discuss the coupling of the radion to matter for the
stabilized setup. A brief discussion on the relevance of our results
for the LRS model and the Mini-Seesaw\footnote{Our results have
  particular relevance for the latter work, as we demonstrate the
  suppressed coupling of KK gravitons and the radion to the UV
  localized SM, and thus prove phenomenological viability of the gravitational sector
  of this model.} appears in Sec.~\ref{sec:applications}, and
Sec.~\ref{sec:ads/cft} attempts to shed light on some of our
findings via the AdS/CFT correspondence. We conclude in
Sec.~\ref{sec:conc} and some technical details are given in an Appendix.
\section{Gravity in a Little Warped Space\label{sec:setup}}
We consider a truncated warped space whose bulk-gravity and UV scales are
much less than the 4D Planck mass, $M_*\ll M_{Pl}$, namely a Little
Warped Space~\cite{McDonald:2010jm,Duerr:2011ks}. To study the effects
of a large UV curvature term on the spectrum of metric fluctuations we
will employ the interval approach to brane-world
gravity~\cite{Lalak:2001fd,Carena:2005gq,Bao:2005bv}. This approach, which can
reproduce standard RS results obtained on an orbifold, enables a
transparent treatment of boundary curvature terms; these simply modify
the form of the boundary Einstein equations, and thus modify the
boundary conditions (BCs) for the metric fluctuations. Before
proceeding, we note that previous works have considered the effects of
brane curvature terms in the RS framework using the orbifold
picture~\cite{Davoudiasl:2003zt,Davoudiasl:2005uu}, and for
$AdS_5/AdS_4$ in the
interval approach~\cite{Carena:2005gq,Bao:2005bv}.

We consider a warped extra
dimension, described by the
coordinate $y\in[0,L]$, such that a UV (IR) brane of characteristic
energy scale $M_*$ ($e^{-kL}M_*$) is located at
$y=0$ ($y=L$). The metric is given by
\beq
ds^2 = e^{-2ky}\eta_{\mu\nu}dx^{\mu}dx^{\nu} +
dy^2= G_{MN} dx^{M}dx^{N},
\label{bulkmetric}
\eeq
where $M,N,..$ ($\mu,\nu,..$) are the 5D (4D) Lorentz indices and
$k$ is the $AdS_5$ curvature. The action for the LWS,
with brane localized curvature terms included, is
\bea
S&=&\int_{\mathcal{M}} d^5x \sqrt{-G}\
\left\{2M_*^3\mathcal{R}-\Lambda\right\}\ +\ \sum_i\int d^4x \sqrt{-g}\left\{M_i^2 R- V_i/2\right\}\nonumber\\
& &\qquad\qquad+\ 4M_*^3\oint_{\partial \mathcal{M}} \sqrt{-g}\ K,\label{gravity_action}
\eea
where $\mathcal{R}$ is the bulk Ricci scalar constructed from $G_{MN}$
and $M_*$ is the 5D gravity scale.  The brane localized curvature $R$ has coefficient $M_i$ on the $i$th boundary ($i=0,L$) and is constructed with $g_{\mu\nu}$ (the restriction of $G_{\mu\nu}$ to the relevant boundary).
The last term is the usual Gibbons-Hawking boundary term~\cite{Gibbons:1976ue},
with $K$ being the extrinsic curvature of the manifold $\mathcal{M}$. This term
is necessary to obtain consistent Einstein equations on an interval~\cite{Lalak:2001fd}.
The gravitational field is sourced by the  bulk
cosmological constant $\Lambda$ and brane tensions $V_i$. The bulk
curvature is $k=\sqrt{-\Lambda/24M_*^3}$ and the brane tensions take their usual RS values,
$V_i=-24kM_*^3\theta_i$, with $\theta_0=-\theta_L=-1$. We will employ
the rescaled dimensionless variables $v_i=M_i^2k/M_*^3$ and $w_i=V_i/2M_*^3k$ for
the brane quantities.

The effective 4D Planck mass has contributions from both the bulk and
brane intrinsic curvatures, and is calculated as
\bea
M_{Pl}^2=\frac{M_*^3}{2k}\left\{1-e^{-2kL}+v_0
  + v_Le^{-2kL}\right\}.
\eea
We will be interested in the limit $M_0\gg M_*$, corresponding to $v_0\gg1$. In this limit the
4D Planck mass is predominantly determined by the coefficient of the
UV Einstein-Hilbert term:
\bea
M_{Pl}^2\simeq \frac{M_*^3}{2k}\times v_0 = \frac{M_0^2}{2}.
\eea
Thus, provided $M_0$ is suitably large, a viable 4D Planck mass can
be obtained even when
the bulk gravity scale is much less than the Planck scale, $M_*\ll
M_{Pl}$. This is the main reason to
introduce the large UV term. In addition to modifying the 4D Planck
mass, this term, in general, modifies  the properties of the metric
fluctuations, including the KK gravitons and the radion. Our goal
is to detail the spectrum of these modes and consider the effects
of the large UV term on the low energy particle (KK) spectrum.

Varying the action gives the bulk equations of motion,
\bea
\mathcal{R}_{MN}-\frac{1}{2}G_{MN}\mathcal{R}=-\frac{\Lambda}{4M_*^3}G_{MN}.
\eea
The boundary conditions are obtained by combining the variations of the
4D brane action and the Gibbons-Hawking term with the surface terms arising
from the variation of the bulk action.  The result is~\cite{Carena:2005gq}
\bea
\left[\frac{v_i}{k}
  \left(R_{\mu\nu}-\frac{1}{2}g_{\mu\nu}R\right)+\frac{1}{2}g_{\mu\nu}kw_i+\theta_i\sqrt{G^{55}}(g_{\mu\nu,5}-g_{\mu\nu}\
  g_{\alpha\beta,5}\
  g^{\alpha\beta})\right]_{y=y_i}=0.
\eea
The notation here is that the above equation must be evaluated
separately at $y=0,L$.
Without loss of generality we work in a
``straight'' gauge defined by
$G_{\mu5}=0$~\cite{Carena:2005gq}. Expanding the metric in terms of a
fluctuation, $G_{MN}= G_{MN}^0+h_{MN}$, where the zeroth order metric
is given by $G_{\mu\nu}^0=e^{-2ky}\eta_{\mu\nu}$, $G_{55}^0=1$ and
$G_{M5}^0=h_{M5}=0$ in a straight gauge, the boundary condition can be
expressed as
\bea
& &\left[\frac{v_i}{2k}\left\{h_{\alpha\mu,\nu}^{\ \ \ \ \alpha}
    +h_{\alpha\nu,\mu}^{\ \ \ \ \alpha}-h_{\mu\nu,\alpha}^{\ \ \ \ \alpha}-\tilde{h}_{,\mu\nu}-g_{\mu\nu}(h_{\alpha\beta,}^{\ \ \
      \ \alpha\beta}-\tilde{h}_{,\alpha}^{\ \alpha})\right\}\right.\nonumber\\
& &\qquad \qquad+\ \left. \theta_i\left\{2kh_{\mu\nu}+h_{\mu\nu,5}-g_{\mu\nu}\tilde{h}_{,5}-3kg_{\mu\nu}h_{55}\right\}\frac{}{}\right]_{y=y_i}=0,
\eea
where indices are raised with $g^{\mu\nu}=e^{2ky}\eta^{\mu\nu}$.

In a straight gauge we can always use remnant gauge freedom to write
the metric fluctuation $h_{55}$ as~\cite{Carena:2005gq,Bao:2005bv}
\bea
h_{55}(x^\mu,y)=F(y) \psi(x^\mu),\label{h_55_fluctuation}
\eea
where $F(y)$ is an arbitrary function of $y$ satisfying $\int_0^Ldy\
F(y)\ne0$. One can always recast an arbitrary $h_{55}$ into the form
(\ref{h_55_fluctuation}) by performing a general 5D coordinate
transformation $x^M\rightarrow x^M+\xi^M$ with $\xi^\mu=0$ and~\cite{Carena:2005gq,Bao:2005bv}
\bea
\xi^5=\frac{1}{2}\int^ydy\, h_{55} -\frac{1}{2}\int^ydy\, F(y) \psi.
\eea
The ability to specify an arbitrary $F(y)$ is a remnant gauge freedom.

For massive 4D modes the tensor $h_{\mu\nu}$ can be written as
\bea
h_{\mu\nu}\rightarrow h_{\mu\nu}+\partial_\mu V_\nu +\partial_\nu
V_\mu
+e^{-2ky}\partial_\mu\partial_\nu\mathcal{S}_1+G^0_{\mu\nu}\mathcal{S}_2,
\eea
where $h_{\mu\nu}$ is now transverse-traceless with five degrees of freedom, $\partial^\alpha
h_{\alpha\beta}=\eta^{\alpha\beta}h_{\alpha\beta}=0$, and $V_\mu$ is
transverse, $\partial^\alpha V_\alpha=0$. $\mathcal{S}_1$ and
$\mathcal{S}_2$ are scalar degrees of freedom. One can show that the
physical massive modes are contained in $h_{\mu\nu}$. This has been
discussed in detail in~\cite{Carena:2005gq,Bao:2005bv} and we have
verified their results. Here we simply note a few key points
(using the notations of~\cite{Carena:2005gq} so the reader can readily
find more details there). One can use the bulk
equations of motion to show that $\partial_5(e^{2ky} \bar{V}_\mu)=0$,
which fixes the $y$-dependence of $V_\mu$ as $e^{-2ky}$ ($\bar{V}_\mu$ is the
Fourier transform of $V_\mu$). Thus one can use the transverse part of
the remaining $\xi^\mu$ gauge freedom to remove $V_\mu$. The
longitudinal part of the $\xi^\mu$ gauge freedom can be used to remove
one scalar degree of freedom so that $\mathcal{S}_1$ and
$\mathcal{S}_2$ can be expressed in terms of $\psi$ and an unspecified
function $f_1(x)$. For massive modes the boundary conditions can be
shown
to require\footnote{The ``bars'' again denote 4D Fourier transforms.}
\bea
 \left(\theta_i-v_i\right)\times\left[\frac{e^{-2ky}\bar{f}_1}{\eta^{\mu\nu}p_\mu p_\nu}+\mathcal{F}\bar{\psi}\right]_{y=y_i}=0,
\eea
where $\mathcal{F}'= F$. Thus the boundary conditions force $\bar{f}_1=\bar{\psi}=0$,
assuming the brane-curvature coefficients are not tuned to satisfy
$\theta_i= v_i$. Together, this shows that the only 
physical, massive degrees of
freedom are contained in $h_{\mu\nu}$. 

KK expanding the physical fluctuations as
\bea
h_{\mu\nu}(x,y)=\kappa_*\sum_n h^{(n)}_{\mu\nu}(x)f_n(y),
\eea
where $\kappa_*$ is chosen to give the 4D fields
$h_{\mu\nu}^{(n)}$ a canonical mass dimension, the bulk equations of motion reduce to
\bea
(\partial_5^2 +e^{2ky}m_n^2-4k^2)f_n(y)=0,
\eea
giving
\bea
f_n(y)=\frac{1}{N_n}\left\{J_2\left(\frac{m_n}{k}e^{ky}\right)
+\beta_n Y_2\left(\frac{m_n}{k}e^{ky}\right)\right\}.
\eea
Here, $m_n$ is the mass of the $n^\text{th}$ KK mode.
The boundary conditions require
\bea
\left.\left[\partial_5 +2k-e^{2ky}m_n^2
    \frac{\lambda_i}{2\theta_i}\right]f_n(y)\right|_{y=y_i}=0,
\eea
enforcing which gives
\bea
\beta_n=-\frac{ J_1(z_i)
  -z_iv_i\theta_i/2J_2(z_i)}{ Y_1(z_i)
  -z_iv_i\theta_i/2Y_2(z_i)},\label{massive_KK_beta}
\eea
with $z_i=m_ne^{ky_i}/k$. Correcting for the different definitions of
the parameters in
the action, this result agrees with that obtained via the orbifold
approach in Ref.~\cite{Davoudiasl:2003zt} (see Eq.~(2.9) in the
published version).  

The KK spectrum is found by equating the values of $\beta_n$ obtained at the distinct boundaries. For comparison, note that the usual
KK graviton spectrum in RS models is determined by~\cite{Davoudiasl:1999jd}
\bea
\beta_n^{RS}= -\frac{J_1(z_0)}{Y_1(z_0)}=-\frac{J_1(z_L)}{Y_1(z_L)},
\eea
corresponding to Neumann BCs at both the IR and UV branes. The usual
RS KK masses are therefore (approximately) determined by the solutions
to $J_1(m_n^{RS}
e^{kL}/k)=0$.
In the limit of a large UV-brane curvature term, $v_0\gg
1$, with $v_L=0$ for simplicity, Eq.~(\ref{massive_KK_beta}) gives
\bea
\beta_n= -\frac{J_1(z_L)}{Y_1(z_L)} \simeq
-\frac{J_2(z_0)}{Y_2(z_0)}.\label{LWS_KK_masses}
\eea
The UV BC is changed from Neumann to (approximately) Dirichlet for
$v_0\gg1$, while the IR BC is unchanged.\footnote{This is expected
for $v_L=0$. Note that $v_L\ne 0$ will modify the IR BC~\cite{Davoudiasl:2003zt}, but not to
the extent that the UV BC is modified by $v_0\gg1$, due to the
constraint $v_L<1$ that we obtain in Sec.~\ref{sec:_massless_radion}.}  For KK modes with mass $m_n/k\ll 1$ the spectrum
remains determined by the solutions to $J_1(m_ne^{kL}/k)=0$. Light KK modes
are localized towards the IR brane and are not significantly affected
by the large UV term. The spectrum becomes modified for larger $n$
such that $m_n/k$
approaches unity; in the absence of a UV brane curvature term, heavier modes have more significant overlap with
the UV brane and are therefore
more affected by this term, which tends to repel them from the
UV brane.

Now we compute the normalization factors $N_n$.
Taking $\kappa_*^2=1/(2M_*^3)$ we have
\bea
N_n^2 &=& \int dy e^{2ky} \left\{J_2\left(\frac{m_n}{k}e^{ky}\right)
+\beta_n Y_2\left(\frac{m_n}{k}e^{ky}\right)\right\}^2 \nonumber\\
& &\quad+  \sum_i \frac{v_i}{2k}\left\{J_2\left(\frac{m_n}{k}e^{ky_i}\right)
+\beta_n Y_2\left(\frac{m_n}{k}e^{ky_i}\right)\right\}^2.
\eea
In the LWS limit of $v_0\gg 1$, with $v_L=0$ again for simplicity, one has
\bea
N_n&\simeq&\frac{e^{kL}}{\sqrt{2k}}|J_2(z_L)|\left[1+\frac{e^{-2kL}}{|J_2(z_L)|^2}\frac{1}{v_0}\right]^{1/2}\simeq \frac{e^{kL}}{\sqrt{2k}}|J_2(z_L)|.
\eea
With this result one can consider the UV coupling of KK modes, the strength of which is sensitive to the UV value of the wavefunction:\footnote{The UV coupling of KK gravitons has previously been considered in the context of DGP gravity with a flat infinite extra dimension~\cite{Dvali:2001gm,Dvali:2001gx}.}
\bea
f_n(y=0)\simeq
-\frac{\sqrt{2k}}{e^{kL}}\frac{1}{|J_2(z_L)|}\frac{1}{v_0}\simeq
-\frac{\sqrt{\pi }}{e^{kL/2}}\frac{\sqrt{m_n}}{v_0},
\eea
where we have made use of (\ref{LWS_KK_masses}) for $m_n/k\ll1$. Observe that the UV value is suppressed like $1/v_0\ll1$ and,
in particular, one has $f_n(0)=0$ for
$v_0\rightarrow\infty$. Thus, in the limit $M_{Pl}\rightarrow\infty$, the KK gravitons are repelled from the UV brane. Despite this repulsion,
the KK modes remain in the spectrum for
$v_0\rightarrow\infty$.

The spectrum of metric fluctuations also contains a massless spin-two mode and a massless scalar mode. We will discuss the scalar mode in detail in the next
section. The profile for the spin-two zero mode is
\bea
f_0(y)=e^{-2ky}\sqrt{ \frac{2k}{ 1-e^{-2kL}+\sum_i v_ie^{-2ky_i}}},
\eea
and, with this wavefunction, the zero mode is identified as the massless
4D graviton. This mode couples with strength
$M_{Pl}^{-1}$ and reproduces Einstein gravity. This coupling
strength does not depend on the localization of the source field
$T_{\mu\nu}$ and, to leading order, Einstein gravity is recovered for
all fields in the warped space. Note that in the limit $v_0
\rightarrow \infty$ one has $M_{Pl}\rightarrow \infty$ and
$f_0(y)\rightarrow0$. Thus, in this limit the zero mode graviton is
removed from the spectrum and 4D gravity is decoupled from the
theory.

\section{The Little Radion\label{sec:_massless_radion}}
As mentioned already, the spectrum of metric fluctuations also
contains a massless scalar mode, namely the
radion. In physical theories the radion must be massive if the length
of the extra dimension is to be stabilized. We will detail an
implementation of the Goldberger-Wise mechanism for stabilizing the LWS in
Sec.~\ref{sec:little_GW}. However, it will first prove instructive
to consider
the massless radion. The radion only acquires its mass after the
backreaction of the stabilizing dynamics is
included~\cite{Csaki:2000zn}, so the results obtained below for a
massless radion will, in some instances, provide a good approximation for the calculable
case of a weak backreaction. 

We find it convenient to
write the most general form of the metric, including background and scalar perturbations, as
\bea
G_{MN}=\begin{pmatrix}
    a^2\left[\eta_{\mu\nu}
        +\nabla_\mu\nabla_\nu P_3
        +\eta_{\mu\nu}\left(2P_2-aa'P_3'\right)\right] & 0 \\
    0 & 1+2P_1-\left(a^2P_3'\right)'
\end{pmatrix} \:,
\label{p_metric_ansatz}
\eea
where $a(y)$ is the background warp factor, and $P_1$, $P_2$ and $P_3$
are spin-zero perturbations that depend on $x^\mu$ and $y$.  This form
of
the metric is
inspired by the gauge-invariant forms discussed
in~\cite{Bridgman:2001mc,Deffayet:2002fn}. With this parameterization,  the Einstein
equations have a particularly simple structure. Two of
the bulk Einstein equations for the perturbations can be taken as
\bea
\partial_\mu\partial_\nu\left(P_1+2P_2\right) &=& 0 \qquad  \mu\ne\nu \:,\label{einstein_relate_p1p2}\\
\partial_\mu\left(\frac{a'}{a}P_1 - P_2'\right) &=& 0 \qquad  \forall\mu\:.
\eea
Taking the integration ``constants'' to be zero,\footnote{As required
  by the demand that all
  perturbations be localized in $x$.}  the first equation
relates $P_2$ to $P_1$, while the second determines the $y$-dependence of $P_1$.  Using these results, the remaining bulk Einstein
equation is simply $\Box P_1=0$, which is satisfied for a massless 4D field.
The perturbation $P_3$ is completely free in the bulk  and is a
remnant gauge freedom.\footnote{See Appendix~\ref{app:gauge} for a discussion of the remaining gauge freedom in our chosen straight gauge.} This is a
generalization of the remnant gauge freedom in the massless sector described
by the arbitrary function $F(y)$ in~\cite{Carena:2005gq,Bao:2005bv},
and physical quantities do not depend on $P_3$. The boundary
conditions derive from the two additional
boundary Einstein equations:
\bea
P_3'(y_i)=\frac{-v_i}{a(y_i)\left[\theta_ika(y_i)+v_ia'(y_i)\right]} P_1(y_i) \:.
\label{p1p3_boundary_condition}
\eea

Using the solutions to the above we can compute the effective 4D
action for the physical scalar fluctuation.  We
perform separation of variables and solve for the profile of $P_1$,
giving
\bea
P_1= a^{-2}(y)\psi(x^\mu) \: \label{p1_expression}.
\eea
This solution is consistent with the boundary
conditions~\eqref{p1p3_boundary_condition} for general $v_i$
so long as one chooses a $P_3$ with $P_3'\ne 0$ at the boundaries.
For the sources given in (\ref{gravity_action}), the solution for the background
metric is of the standard RS form, $a(y)=e^{-ky}$.
Ignoring the 4D surface terms ($\psi$ vanishes at
$x^\mu\rightarrow\infty$), one can derive the action for the scalar
perturbations up to quadratic order in $\psi$ as
\bea
\mathcal{S}_{\mathcal{O}(\psi^2)} = \int d^4x \left[
    \frac{3M_*^3}{k} e^{2kL}
    \left(\frac{1}{1-v_L} - \frac{e^{-2kL}}{1+v_0}\right)
  \right]
  \left( -\frac12 \eta^{\mu\nu} \partial_\mu \psi \partial_\nu \psi \right) \:\label{radion_kinetic_GI}.
\eea
 It is worth noting that there are no terms linear in
$\psi$, or additional quadratic terms, in the action
(\ref{radion_kinetic_GI}). This provides an important check on the
consistency of the calculation. In particular, the mass
terms\footnote{The radion will acquire mass once we specify
  a mechanism to stabilize the length of the extra dimension.} and
higher-order derivative terms which appear at the quadratic level in
individual terms in the action (\ref{gravity_action}) cancel out in
the full action. It is essential to include the Gibbons-Hawking term
in the action to achieve this result.

The physical radion is defined as $r(x)=\psi(x)/N_\psi$, where
 $N_\psi$ is the following normalization constant:
\bea
N_\psi^2=\frac{k}{3M_*^3}e^{-2kL}\frac{(1-v_L)(1+v_0)}{(1+v_0)-(1-v_L)e^{-2kL}}.
\eea
We see that the kinetic term is only well behaved for $v_L<1$, giving
an upper bound on the size of the IR localized brane kinetic term. The UV term suffers
no such constraint and one may safely take $v_0\gg1$. In this limit
one has
\bea
N_\psi^2\simeq\frac{k}{3M_*^3}e^{-2kL}(1-v_L)\left[1+\frac{(1-v_L)e^{-2kL}}{v_0}\right],\qquad v_0\gg1\ ,
\eea
and, to leading order, the dependence on $v_0$ drops out .
\section{Little Radion Coupling to Matter\label{sec:radion-matter-GI}}
In the previous section we detailed the physical gravi-scalar in the LWS (the  ``little radion''~\cite{Davoudiasl:2010fb}). We would now like to determine the coupling of this mode to localized matter. The large UV curvature term modifies the wavefunction of the radion and is expected to alter its couplings. We will detail the dependence of the couplings on the UV term in this section. Our results will reduce to standard RS results in the limit $v_i\rightarrow0$, allowing us to contrast the coupling of the little radion to the standard case with $v_i=0$.  

The coupling of the radion to brane matter depends on the localization of the matter. To demonstrate our points we consider a scalar localized on the boundary at $y=\yp$ with
action 
\bea
S_\phi=-\frac{1}{2}\int d^5x
\sqrt{-g}\left\{g^{\mu\nu}\partial_\mu\phi\partial_\nu\phi
  +V(\phi)\right\}\delta(y-y_\phi).\label{brane_scalar_action_rs}
\eea
First
consider the propagation of $\phi$ in the background metric. After
integrating out the extra dimension the action is cast as
\bea
S_\phi=-\frac{1}{2}\int d^4x
\left\{\eta^{\mu\nu}\partial_\mu\phi\partial_\nu\phi
  +V(\phi)\right\},\label{brane_scalar_action_GI}
\eea
where we have rescaled $e^{-k\yp}\phi\rightarrow\phi$ to bring the
kinetic term to a canonical form. Thus the dimensionful parameters in the
potential $V(\phi)$ are ``warped down'' parameters, e.g.\ the mass in
$V(\phi)$ is $m_\phi=e^{-k\yp}m_0$, where $m_0$ is the bare input mass
parameter.

Now return to the original action (\ref{brane_scalar_action_rs}) and expand the metric in terms of the scalar fluctuation as
$g_{\mu\nu}\rightarrow g_{\mu\nu}+f_{\mu\nu}$, where $f_{\mu\nu}$
contains only the spin-zero parts of the perturbation. Integrating over
the extra dimension, the action
becomes
\bea
S_\phi&=&S_0-\frac{1}{2}\int
d^4x\sqrt{-g}f^{\mu\nu}\tilde{T}_{\mu\nu}\nonumber\\
& &\quad-\frac{1}{8}\int
d^4x\sqrt{-g}\left\{-f^\mu_\nu f^\nu_\mu +\frac{1}{2}(f^\mu_\mu)^2\right\}\left\{g^{\mu\nu}\partial_\mu\phi\partial_\nu\phi
  +V(\phi)\right\}\nonumber\\
& &\quad 
+\int
d^4x\sqrt{-g}\left(f^\mu_\alpha f^{\alpha\nu}
-\frac{1}{2}f^{\mu\nu}f^\alpha_\alpha\right)\partial_\mu\phi\partial_\nu\phi +\dots
\eea
where
\bea
\tilde{T}_{\mu\nu}= -\partial_\mu\phi\partial_\nu\phi+\frac{1}{2}g_{\mu\nu}\left\{g^{\alpha\beta}\partial_\alpha\phi\partial_\beta\phi
  +V(\phi)\right\}.\label{stress_energy_scalar}
\eea
Rescaling $\phi$ to bring $S_0$ to a canonical form, we can write the
term linear in the fluctuation as
\bea
\left.S_{\phi}\right|_{\mathcal{O}(f)}=-\frac{1}{2}e^{2k\yp} \int
d^4x\ 
\eta^{\mu\alpha}\eta^{\nu\beta}f_{\mu\nu}\ T_{\alpha\beta} \:, \label{fluc_coupling_T}
\eea
where $T_{\mu\nu}$ is now written in terms of canonical fields and the flat space metric; i.e.\ replace $g_{\mu\nu}\rightarrow \eta_{\mu\nu}$ in (\ref{stress_energy_scalar}) to obtain $T_{\mu\nu}$.

The gravi-scalar couples to $T_{\mu\nu}$ both with and without
derivatives. Let us consider the non-derivative couplings:
\bea
\left.S_{\phi}\right|_{\mathcal{O}(f)}=\frac{e^{2ky_\phi}}{2}\left[1-\frac{v_ia'}{a(\theta_ika+v_ia')}\right]\int
d^4x\ \psi\ T+\dots \label{S_f_general gauge_GI}.
\eea
 where $T=\eta^{\mu\nu}T_{\mu\nu}$ and we have made use of
Eqs.~(\ref{einstein_relate_p1p2})-(\ref{p1_expression}). We are interested primarily in the effects of the large UV term, so
 let us simply set $v_L=0$. Then the coupling of the physical radion
 is 
\bea
\left.S_{\phi}\right|_{\mathcal{O}(f)}=\frac{1}{2} \int
d^4x\ \frac{r}{\Lambda_{r}}\times T+\dots\ ,\label{S_f_L gauge_GI}
\eea
where the coupling is controlled by the dimensionful parameter
\bea
\Lambda_{r}^{-1}&=& \sqrt{\frac{k}{3M_*^3}}\sqrt{\frac{1+v_0}{1+v_0-e^{-2kL}}}\times \left\{
\begin{array}{lc}
e^{-kL}/(1+v_0) &\quad\yp=0\\e^{kL} &\quad\yp=L
\end{array}
\right. .
\eea

Let us consider two limits of the above expression. First, consider the RS limit of $v_0\rightarrow0$ with $M_*\sim M_{Pl}$:
\bea
\left.\Lambda_{r}^{-1}\right|_{\mathrm{RS}}&\simeq& \sqrt{\frac{k}{3M_*^3}} \times \left\{
\begin{array}{lc}
e^{-kL}&\yp=0\\e^{kL} &\yp=L
\end{array}
\right. .
\eea
Observe that the RS radion couples to UV localized fields with
strength $\sim e^{-kL}/M_*\sim e^{-kL}/M_{Pl}$ and IR localized fields with
strength $\sim 1/(e^{-kL}M_{Pl})\sim e^{kL}/M_*$ (taking $k\sim M_*$
in both cases). The
coupling to UV (IR) fields is thus suppressed (enhanced) relative to
the bulk gravity scale $M_*$. Note that the IR coupling demonstrates the
standard result that the radion couples to IR fields with a strength set
by the inverse IR scale $(e^{-kL} M_*)^{-1}$~\cite{Goldberger:1999un,Csaki:2000zn,Cheung:2000rw}.  

Next consider the
LWS limit $M_*\ll M_{Pl}$, which requires $v_0\gg 1$ to achieve a
viable 4D Planck mass:
\bea
\left.\Lambda_{r}^{-1}\right|_{\mathrm{LWS}}&\simeq& \sqrt{\frac{k}{3M_*^3}} \times \left\{
\begin{array}{lc}
e^{-kL}/v_0&\yp=0\\e^{kL} &\yp=L
\end{array}
\right. .
\eea
Observe that the coupling to IR fields is the same as that obtained in
the RS limit, $\Lambda_r\sim e^{-kL}M_*$. The large UV term does not modify the IR-brane coupling
of the radion to leading order. For UV localized fields the radion
coupling picks up a factor of $v_0^{-1}\ll1$ and  is
significantly suppressed relative to the UV scale $M_*$ in the LWS
limit. Note, however, that when expressed in terms of the 4D Planck
mass, the coupling of the radion to UV matter remains as $\Lambda_r\sim
e^{kL}M_{Pl}$. Also observe that in the limit $v_0\rightarrow \infty$
the radion is repelled from the UV brane, while the IR coupling is
unaffected. We will return to this feature in subsequent discussions.

Let us note that in the limit $v_i\rightarrow0$ one can use the
remaining gauge freedom to choose the form of the scalar fluctuations
such that the derivative pieces in~\eqref{p_metric_ansatz},
$\nabla_\mu\nabla_\nu P_3$ and $P_3'$, vanish.
Thus, in the limit that the brane curvature terms are turned off, the standard
parametrization of the scalar metric fluctuations in RS~\cite{Goldberger:1999un,Csaki:2000zn,Cheung:2000rw} is consistent
with the boundary conditions. However, for $v_i\ne0$ we find that one
cannot remove the derivative pieces in~\eqref{p_metric_ansatz} with a
gauge choice and \emph{simultaneously} obtain a solution that is consistent with the
boundary conditions. We have checked this result using the
metric parametrization of Refs.~\cite{Carena:2005gq,Bao:2005bv} and
arrive at the same conclusion. 
\section{Stabilizing a Little Radius\label{sec:little_GW}}
In the preceding sections we have detailed the spectrum of metric
fluctuations in a LWS. In that analysis, however, no stabilizing
dynamics were introduced so that the size of the extra dimension was
not fixed and the radion remained
massless. Realistic physical theories constructed on a LWS, like the
Little RS model and the Mini-Seesaw, 
require the radion to develop a mass. Thus we must consider the
effects of some stabilizing dynamics if we are to accurately report the
spectrum of metric fluctuations in physically realistic cases.

We will follow the original idea of Goldberger and
Wise~\cite{Goldberger:1999uk} (also see~\cite{Garriga:2002vf}) and
stabilize the little warped space using a bulk scalar $\Phi$. This produces a KK tower of physical scalars that contain an admixture of the KK modes of $\Phi$ and the gravi-scalar. 
As in the previous sections, we work with the interval approach~\cite{Carena:2005gq,Bao:2005bv}.
Essentially our goal in this section is to generalize the analysis of~\cite{Carena:2005gq}
to include a 5D scalar with bulk and brane potentials.
The complete action is therefore
\begin{align}
\mathcal{S} &=
  \int_{\mathcal{M}} d^5x \sqrt{-G}
    \left\{2M_*^3 \mathcal{R}
      - \frac12 G^{MN}\partial_M\Phi\partial_N\Phi - V(\Phi)\right\} \nonumber\\
&\qquad
  + \sum_i \int d^4x \sqrt{-g}
    \left\{\frac{M_*^3v_i}{k} R - M_*^3k w_i
      - \frac14 t_i g^{\mu\nu} \partial_\mu\Phi\partial_\nu\Phi - \frac12 \lambda_i(\Phi)\right\} \nonumber\\
&\qquad
  + 4M_*^3 \oint_{\partial \mathcal{M}} \sqrt{-g}\, K \:.
\label{5d_stabilized_action}
\end{align}
As before, $G_{MN}$ is the bulk metric and $g_{\mu\nu}$ the induced 4D boundary metric,
with corresponding Ricci scalars $\mathcal{R}$ and $R$.
We have allowed for brane kinetic terms in both the gravity
($v_i$) and scalar ($t_i$) sectors.  $V$ is the bulk potential for the
scalar $\Phi$, which now subsumes the bulk cosmological constant, and
$\lambda_i$ are the brane localized potentials.
The brane tensions $kw_i$ are explicitly separated from the brane
potentials such that $\lambda_i(\Phi)=0$ for the background value of $\Phi$.

 Let us emphasize that, in this section, we are not specifying the
 form of the background metric. The point is that the potentials
 $V(\Phi)$ and $\lambda_i(\Phi)$ will cause $\Phi$ to obtain a nonzero
 background value which, along with the bulk cosmological constant and
 the brane tensions, will source the metric. The analysis of this
 section will be performed for arbitrary potentials $V(\Phi)$ and
 $\lambda_i(\Phi)$, subject to the constraint that the metric
 preserves a 4D Lorentz symmetry. Ultimately we are interested in the
 case where the metric is a perturbed version of the standard RS form,
 in order to determine the mass of the little radion and to
 demonstrate stabilization in the LWS. We will specify a background
 solution appropriate for this case in the next subsection. However, our results in this section are completely general. 

As usual, varying the above action with respect to the degrees of
freedom gives the equations of motion. Varying with respect to the metric gives the bulk Einstein
equations
\bea
  \left(\mathcal{R}_{MN}-\frac12G_{MN}\mathcal{R}\right)
    - \frac{1}{4M_*^3}\left(
      \partial_M\Phi\partial_N\Phi
      - \frac12 G_{MN} G^{PQ} \partial_P\Phi\partial_Q\Phi
      - G_{MN} V
    \right) = 0 \:,
\eea
and the boundary Einstein equations
\begin{align}
&
\bigg[
  \frac{v_i}{k}\left(
    R_{\mu\nu}
    - \frac12g_{\mu\nu}R
  \right)
    + \frac12g_{\mu\nu}kw_i
    + \theta_i\sqrt{G^{55}}\left(
      g_{\mu\nu,5}
      - g_{\mu\nu}\,g_{\alpha\beta,5}\,g^{\alpha\beta}
    \right)
\nonumber\\&\qquad
    - \frac{1}{4M_*^3}\left(
      t_i\partial_\mu\Phi\partial_\nu\Phi
      - \frac12t_ig_{\mu\nu}\,g^{\alpha\beta}\partial_\alpha\Phi\partial_\beta\Phi
      - g_{\mu\nu}\lambda_i
    \right)
\bigg]_{y=y_i} = 0 \:.
\end{align}
Variation with respect to the field $\Phi$ gives the bulk
Euler-Lagrange equation
\bea
  \partial_M\left(\sqrt{-G}\,G^{MN}\partial_N\Phi\right)
    - \sqrt{-G}\,V_{,\Phi} = 0 \:,
\eea
where the subscript $\Phi$ on $V$ denotes a derivative.  The
corresponding boundary equations are
\bea
\left[
  t_i\partial_\mu(\sqrt{-g}\,g^{\mu\nu}\partial_\nu\Phi)
    - \sqrt{-g}\,\lambda_{i,\Phi}
    - 2\theta_i\sqrt{-G}\,G^{5N}\partial_N\Phi
\right]_{y=y_i} = 0 \:.
\eea
Note that these boundary equations are expressed in straight gauge.

Our aim now is three-fold.  We first solve for the background configuration
of the above equations.  This is straightforward; our solutions
will be generic and expressed as a set of differential equations to
be solved for a specific scalar potential.  Following this we solve for
first order spin-zero perturbations around the background, giving us the
spectrum of scalar KK excitations.  The answer will be in the form of a
Schr\"odinger equation, whose potential depends on the background
configuration from the first step.  Third, we expand the original
action to second order in the perturbations (expanded around the
background) in order to compute the normalization condition for the
KK modes.

Taking the usual warped metric ansatz:
\bea
ds^2=a^2(y)\eta_{\mu\nu}dx^\mu dx^\nu +dy^2,
\eea
where the warp factor $a(y)$ is to be determined, and allowing the background value
of $\Phi$ to depend
only on $y$ [denoted as $\phi(y)$], the background is solved by
\bea
0 &=& \frac{a''}{a} - \frac{a'^2}{a^2} + \frac{\phi'^2}{12M_*^3} \:,\\
0 &=& 24M_*^3\frac{a'^2}{a^2} - \frac12\phi'^2 + V(\phi) \:,\\
0 &=& \left[ w_ik - 12\theta_i\frac{a'}{ka} \right]_{y=y_i} \:,\\
0 &=& \left[ \lambda_{i,\Phi}(\phi) + 2\theta_i\phi' \right]_{y=y_i} \:.
\eea
The first two equations relate bulk quantities, while the second two
are required to satisfy the boundary equations.  In this case, the
fixed value of $w_i$ is the usual fine tuning for the RS brane tensions.

For the perturbations we use the same metric ansatz as in the non-stabilized
case, Eq.~\eqref{p_metric_ansatz}, along with the scalar perturbation
\bea
\Phi(x^\mu,y)=\phi(y)+P_4(x^\mu,y) \:.
\label{phi_perturbation}
\eea
The $(\mu,\nu)$ and $(\mu,5)$ bulk Einstein
equations give
\bea
\partial_\mu\partial_\nu(P_1+2P_2) &=& 0 \qquad \mu\ne\nu \:,\\
\partial_\mu\left(
  \frac{a'}{a}P_1
  -P_2'
  -\frac{a^2\phi'^2}{24M_*^3}P_3'
  -\frac{\phi'}{12M_*^3} P_4 \right) &=& 0 \qquad \forall \mu \:.
\eea
The integration constants must be zero, similar to the non-stabilized
case, allowing us to use the first equation to solve for $P_2$ and the
second to solve for $P_4$.  With these solutions, the remaining bulk
equations reduce to a single equation for $P_1$, being
\bea
  -P_1''
    + \left(-2\frac{a'}{a}+2\frac{\phi''}{\phi'}\right) P_1'
    + \left(4\frac{a'}{a}\frac{\phi''}{\phi'}+\frac{\phi'^2}{3M_*^3}\right) P_1
    = \frac{1}{a^2}\Box P_1 \:.
\label{p1_differential_equation}
\eea
As in the non-stabilized case, the function $P_3$ is completely free
in the bulk.  There are four boundary equations for the perturbations.
The first two are the same as in the non-stabilized case,
Eq.~\eqref{p1p3_boundary_condition}.  The additional two result from
the Euler-Lagrange boundary equations and are
\begin{equation}
\begin{aligned}
\bigg[
  \Box P_1
  - \left(
    \frac{2\theta_ia^2\phi''}{\phi'}
    + a^2 \lambda_{i,\Phi\Phi}(\phi)
    - \Box t_i
  \right)
  \left(
    \frac{\theta_i (2a'P_1+aP_1')}{2a}
    + \frac{v_ia\phi'^2P_1}{24M_*^3(ka+\theta_iv_ia')}
  \right)
\bigg]_{y=y_i} = 0 \:.
\end{aligned}
\label{p1_boundary_condition}
\end{equation}
In deriving this boundary condition we have used
Eq.~\eqref{p1_differential_equation} evaluated at the boundaries
(technically, at $y=\epsilon$ and $y=L-\epsilon$ for vanishing $\epsilon$).

To find solutions for $P_1$ we separate variables
\bea
P_1(x^\mu,y)=p_1(y)\psi(x^\mu) \:,
\eea
with $\Box\psi=m^2\psi$.  Then, with an appropriate change of
coordinates and rescaling of $p_1$, Eq.~\eqref{p1_differential_equation}
becomes a Schr\"odinger equation. \footnote{For the extension to
multiple bulk scalars see Ref.~\cite{Aybat:2010sn}.}  This can be solved, along with
the boundary conditions, to find the KK profiles $p_1(y)$ of the
spin-zero perturbations, along with their corresponding mass eigenvalues
$m^2$.  In this way one obtains the physical spectrum of the theory.

We now turn to our third task, which is to compute the effective 4D
action and find the normalization of each mode of the KK tower.
Taking the original 5D action~\eqref{5d_stabilized_action}, we
substitute in the metric ansatz~\eqref{p_metric_ansatz}, with
separation of variables for $P_1$, and use the perturbation ansatz
for $\Phi$~\eqref{phi_perturbation}.  We treat the action order by
order in the perturbations, up to second order.
Our answer takes the form
\bea
\mathcal{S} = \int d^4x \left( \mathcal{L}^{(0)} + \mathcal{L}^{(1)} + \mathcal{L}^{(2)} \right) \:.
\eea
At each order the 4D Lagrangian density has a bulk piece, for which
we shall attempt to do the $y$-integral, and the combined brane and
boundary pieces.

For the zeroth order terms (that is, just the background) the Lagrangian
is
\bea
\mathcal{L}^{(0)} =
  \int_0^L \left(
    -16M_*^3 a^2 a'^2 + \frac13 a^4 \phi'^2
  \right) dy
  +4M_*^3 \sum_i \theta_i a(y_i)^3a'(y_i) \:.
\eea
The first term comes from the bulk piece of~\eqref{5d_stabilized_action}
while the second is the combination of the brane and extrinsic curvature
pieces.  We have used the background equations to obtain this simplified
form.  One can show that the above integrand is equal to $-4M_*^3(a^3a')'$,
and so the integral of these bulk terms exactly cancels the contributions
from the brane and boundary pieces, making $\mathcal{L}^{(0)}$ vanish.
This must be so for consistency; our background metric ansatz has Minkowski
4D slices and so there cannot be an effective 4D cosmological constant.

The first order terms reduce to
\begin{align}
\mathcal{L}^{(1)} &=
  M_*^3 \left[
    2\int_0^L a^2 p_1 dy
    + \sum_i \frac{v_i a(y_i)^3 p_1(y_i)}{ka(y_i) + \theta_i v_i a'(y_i)}
  \right]
  \Box \psi
\nonumber\\&\quad
  + \int_0^L\left[
    \left(-8M_*^3 a^2 a'^2 + \frac16 a^4 \phi'^2 \right)
    \Box P_3
  \right] dy
  + 2M_*^3 \sum_i \theta_i a(y_i)^3 a'(y_i) \Box P_3(x^\mu,y_i) \:.
\end{align}
To obtain this expression we have made use of the background equations
as well as the first order equations for the $P_i$, and we have also
done integration by parts (under the $y$-integral) on all terms that
contain $y$-derivatives of $p_1$ and $P_3$.  Since this Lagrangian
appears under a 4D integral, and the perturbations are assumed to
vanish at 4D infinity, the above expression does not contribute to the
effective 4D action.  This is again as expected.

Deriving the second order piece is a difficult task due to the large number
of bulk, brane and boundary terms that must combine or cancel.  The trick
is to use the bulk background and perturbation equations, and perform
integration by parts, to eliminate all derivatives of $a$, $\phi$ and $p_1$
that are second order or higher.  This brings each term to a canonical form,
allowing the surface terms coming from the $y$ integration by parts to
combine or cancel with the brane and boundary terms.  The factor $p_1''$
must be reduced in some cases by using integration by parts, and in other
cases by using the differential Eq.~\eqref{p1_differential_equation}.
One also makes use of the boundary equation for $p_1$ to further simplify
the resulting terms.  By doing this, the second order terms can be brought
into the expected form for a massive spin-zero field
\bea
\mathcal{L}^{(2)} = \mathcal{N} \left(
  -\frac12 \eta^{\mu\nu}\partial_\mu\psi\partial_\nu\psi
  -\frac12 m^2 \psi^2 \right) \:,
\eea
where $m^2$ is the mass eigenvalue of the KK mode, and the normalization
constant is
\begin{align}
\mathcal{N} &=
  6M_*^3 \int_0^L \left(
    a^2 p_1^2
    + 24M_*^3 \frac{a'^2}{\phi'^2} p_1^2
    + 24M_*^3\frac{aa'}{\phi'^2} p_1 p_1'
    + 6M_*^3\frac{a^2}{\phi'^2} p_1'^2
    \right) dy
\nonumber\\&\qquad
  + 3M_*^3 \sum_i \frac{v_i a(y_i)^3 p_1(y_i)^2}{ka(y_i)+\theta_iv_ia'(y_i)}
\nonumber\\&\qquad
  + \frac18 \sum_i t_i\left[
    12M_*^3\theta_i\frac{2a'(y_i)p_1(y_i)+a(y_i)p_1'(y_i)}{\phi'(y_i)}
    +\frac{v_ia(y_i)^2\phi'(y_i)p_1(y_i)}{ka(y_i)+\theta_iv_ia'(y_i)}\right]^2 \:.
\label{stabilised_radion_norm}
\end{align}

\subsection{The Radion Mass}

One can use the preceding results to compute the radion mass
[i.e.\ the lightest KK mode of~\eqref{p1_differential_equation}]
for a given background $\phi(y)$. We are interested in the case where the background geometry is of the standard RS form with a weak (calculable) perturbation due to the backreaction of the scalar field. 
We follow Ref.~\cite{Csaki:2000zn} and choose a perturbed background of the form
\bea
a(y) &=& e^{-ky}\left(1-\frac{l^2}{6}e^{-2uy}\right) \:,\\
\phi(y) &=& 2\sqrt{2}M_*^{3/2}le^{-uy} \:,
\eea
which is valid in the $y\in[0,L]$ region. This corresponds to the
potential $V(\Phi)=(W_{,\Phi})^2/2-W^2/6M_*^3$
with $W(\Phi)=12M_*^3k-u\Phi^2/2$, along with the boundary potentials
\bea
\lambda_i(\Phi)=-\theta_iW(\phi_i)-\theta_iW_{,\Phi}(\phi_i)(\Phi-\phi_i)+\gamma_i(\Phi-\phi_i)^2,
\eea
where $u$, $\phi_i$ and $\gamma_i$ are constants. In terms of the input
parameters, the length of the extra
dimension is now fixed at $L=u^{-1}\log(\phi_0/\phi_L)$, and the weak backreaction
limit is defined by $\kappa_*\phi_i/\sqrt{2}\ll1$. We work to order
$l^2$ in the small parameter $l=\kappa_*\phi_0/\sqrt{2}$ in this
section.\footnote{The expression for $\phi$ is actually correct to all
  orders in $l$.} 
The solution for $p_1(y)$ will be a perturbed
form of the massless solution
\bea
p_1(y) = e^{2ky}\left\{1+l^2f(y)\right\} .
\eea
The bulk equation for $f$ turns out to be the same as the case
without brane curvature terms, namely~\cite{Csaki:2000zn}
\bea
f'' + 2(k+u)f' = \frac43u(u-k)e^{-2uy} - \tilde{m}^2e^{2ky} \:,
\eea
where $m^2=l^2\tilde{m}^2$ (the mass of the radion is on the order of the
correction to the background).  Note that the backreaction must be included to obtain a non-zero radion mass. The solution in the bulk is~\cite{Csaki:2000zn}
\bea
f'(y) = -\frac23u\left(1-\frac{u}{k}\right)e^{-2uy}
  - \tilde{m}^2\frac{1}{4k+2u}e^{2ky}
  + Ae^{-2(k+u)y} \:,
\label{radion_solution}
\eea
where $A$ is an integration constant.

For the boundary condition~\eqref{p1_boundary_condition} we shall
work in the limit of stiff brane potentials, $\lambda_{i,\Phi\Phi}\to\infty$.
Then for our perturbed background we have
\bea
\left[ f' + \frac23ue^{-2uy}
  + \frac23\frac{u^2}{k}e^{-2uy} \frac{\theta_iv_i}{1-\theta_iv_i}
\right]_{y=y_i} = 0 \:,
\eea
Using the solution
for the radion in the bulk, Eq.~\eqref{radion_solution}, we can obtain
the mass of the lightest KK spin-zero state
\bea
m^2=\frac{4l^2}{3}\frac{(2k+u)u^2}{k}
  \left(\frac{1}{1-v_L} - \frac{e^{-2kL}}{1+v_0}\right)
  \left(e^{2(k+u)L}-e^{-2kL}\right)^{-1} \:.\label{little_radion_mass}
\eea
This expression for the radion mass generalizes the result of
Ref.~\cite{Csaki:2000zn} for the case of
non-zero brane curvature terms, $v_i\ne0$. Furthermore,  in the limit
$v_i\rightarrow0$ it reduces to the
result in~\cite{Csaki:2000zn}, as expected. For
$v_L=0$, we observe that the difference between $v_0=0$ and $v_0\to\infty$ is
negligible. Therefore the large UV brane term required to include viable 4D
gravity on a LWS does not significantly affect the mass of the radion
and the LWS is suitably stabilized via the Goldberger-Wise
mechanism. Equation~(\ref{little_radion_mass}) also demonstrates the
sensitivity of the radion mass to an IR localized brane curvature
term. Observe that values of the 
coefficient $v_L$ in the range\footnote{Recall that consistency
  demands $v_L<1$.} $0<v_L<1$ tend to increase the mass of the radion
relative to the standard RS result. 

To determine the coupling of the radion to matter in the stabilized LWS we need
to evaluate the normalization constant $\mathcal{N}$ in
Eq.~\eqref{stabilised_radion_norm}. We find 
\bea
\mathcal{N} =
  \frac{3M_*^3}{k}e^{2kL}
    \left(\frac{1}{1-v_L}-\frac{e^{-2kL}}{1+v_0}\right)
    + \mathcal{O}(l^2) \:,
\label{massive_radion_norm_leading_order}
\eea
which, to leading order, is the same as that calculated in the non-stabilized case,
Eq.~\eqref{radion_kinetic_GI}. With this result one can redo the
calculations of Sec.~\ref{sec:radion-matter-GI} to determine the
coupling of the radion to matter. To leading order the couplings agree
with those obtained in Sec.~\ref{sec:radion-matter-GI}.

\section{Applications\label{sec:applications}}
In this section we briefly discuss some applications of our
results.
\subsection{Little Randall-Sundrum Model}
The little RS model attempts to address the SM gauge-hierarchy problem
by employing a LWS with IR (UV) scale of order TeV
($M_*\sim 10^3$~TeV)~\cite{Davoudiasl:2008hx}. It provides a
candidate UV completion for the SM up to energies of order
$10^3$~TeV. In order to be a completely realistic effective theory for energies $E<M_*$,
the LRS model should include 4D Einstein gravity. Though this was not
considered in previous constructs, our results show that by
introducing a large UV brane curvature term, 4D gravity can be
consistently incorporated into this effective theory
framework. In addition to incorporating 4D Einstein gravity, this
approach modifies 
the wavefunctions
of the metric fluctuations. However, from a phenomenological
perspective, this
modification may be unimportant for the KK gravitons; the modification occurs
in the vicinity of the UV brane, while the KK gravitons are strongly
localized towards the IR. One therefore expects the coupling of KK gravitons to
SM fields to be essentially the same as that
in~\cite{Davoudiasl:2008hx}, where a simple Dirichlet UV BC was
employed for the KK gravitons.

On the other hand, we have seen that the wavefunction of the radion is
also modified by the large UV-brane curvature term. Previous works
have employed the standard RS parametrization for the radion
fluctuation to study the phenomenology of this mode in the LRS
framework~\cite{Davoudiasl:2010fb}. As we have seen,
the modification that leads to a Dirichlet BC for the KK
gravitons also modifies the radion wave function so that, strictly
speaking, consistency demands that the radion's wave function should also be modified in LRS
studies employing a Dirichlet UV BC for the KK gravitons. The question then arises, is this modification important in
terms of the coupling and the phenomenology? The
radion couples like a dilaton and thus couples to the trace of the
stress-energy tensor. For on-shell fields, this trace is (classically)
determined by the dimensionful parameters in the Lagrangian. Thus
the coupling of the radion to on-shell SM fermions and massive gauge bosons
occurs locally on the TeV brane, where the Higgs, which breaks both electroweak symmetry
 and the classical scale invariance of the SM, is
localized. In Sec.~\ref{sec:radion-matter-GI} we saw that the
coupling of the radion to IR localized sources was not modified by the large UV curvature term,
relative to the usual RS result. Thus
we expect the radion coupling to on-shell SM fields to be well
approximated by the standard parametrization, even when gravity is
included.\footnote{A detailed discussion of the radion
  coupling to bulk on-shell fields can be found
  in~\cite{Rizzo:2002pq}.} Using our results, one can show that
the form of the radion coupling to the zero mode of a massless
bulk vector (for example, the photon or gluon) matches the form used
in previous analysis~\cite{Davoudiasl:2010fb}, up to corrections suppressed by $v_0$. 

\subsection{A Mini Seesaw in a Little Warped Space}
A different application of a LWS was presented
in~\cite{McDonald:2010jm,Duerr:2011ks}. In these works, the SM was
localized on the UV brane of a LWS whose UV (IR) scale was
$M_*\sim$~TeV ($e^{-kL}M_*\sim$~GeV), with sterile neutrinos propagating in the
bulk. It was shown that realistic theories of neutrino mass can be
constructed in this framework, without recourse to any high energy
(supra-TeV) scale~\cite{McDonald:2010jm}. Our present results show that, by introducing a UV
localized Einstein-Hilbert term with large coefficient, 4D gravity can be
consistently included in this low-energy framework. Furthermore, the current results clear up some
potential concerns regarding the strength with which the KK gravitons
couple to the UV localized SM. We have seen in
Sec.~\ref{sec:setup} and
Sec.~\ref{sec:radion-matter-GI} that, respectively,
the coupling of the KK gravitons and the radion to UV brane
fields is suppressed by the large UV brane term. In both cases
the coupling to the UV brane goes like $1/v_0\ll1$. Thus we conclude
that once low-energy 4D gravity is included in the model
of~\cite{McDonald:2010jm,Duerr:2011ks}, the couplings of the SM to the
KK gravitons and the radion are negligible small and, in particular, the model does
indeed provide
a consistent, viable, low-energy theory. Furthermore, we note that the suppressed coupling of the radion to the UV localized SM has a potentially interesting consequence. If the radion is lighter than the lightest sterile neutrino, the only available decay channels would be to lighter SM fields. Given the weakness of the direct coupling to the SM, the radion may be sufficiently long lived to provide a dark matter candidate. It would be interesting to consider this possibility in detail.
\section{Comments on AdS/CFT\label{sec:ads/cft}}
Via the AdS/CFT correspondence~\cite{Maldacena:1997re}, RS models are
thought to be dual to strongly coupled 4D theories that are
(approximately) conformal for energies $M_*>E> e^{-kL}
M_*$~\cite{ArkaniHamed:2000ds}. The conformal symmetry is broken
explicitly in the UV by a cutoff (dual to the UV brane) and
spontaneously in the IR (dual to the IR brane). UV (IR) localized
fields in the 5D picture are dual to fundamental (composite) fields in
the 4D theory (roughly; see e.g.\ \cite{Batell:2007jv}). Thus the zero-mode graviton, which is localized towards
the UV brane, is dual to a fundamental 4D graviton that is external to
the CFT. The massive KK-gravitons and the radion are, respectively,
dual to spin-two composites and a spin-zero composite dilaton. From
the perspective of the dual 4D theory, a number of features detailed
above are readily understood.

In standard RS models the warping generates the weak/Planck hierarchy
and the UV scale is of order the Planck scale;  the UV scale is thus
associated with the need to cutoff the CFT to include 4D Einstein
gravity. In the absence of a UV localized Einstein-Hilbert term, the
4D Planck mass is generated entirely by the dynamics of the (cut-off)
CFT. That is, the usual expression for the Planck mass in RS: 
\bea
M_{Pl}^2\sim \frac{M_*^3}{k} 
\eea
is dual to an induced Planck mass that  results from
loops\footnote{Strictly, these may be tree-level diagrams with
  external graviton legs, rather than loops.} containing CFT
modes~\cite{Agashe:2002jx}. From the perspective of the dual 4D theory
there is no particular reason why the 4D Planck mass should be
\emph{entirely} induced by the CFT and, more generally, the 4D Planck
mass could contain a ``bare'' contribution that arises either from
integrating out heavy fields present in the UV completion or as a
fundamental input in the theory. Irrespective of the
details of its origin, a bare contribution to the
Planck mass in the 4D theory is consistent with the symmetries of the
theory and, in the standard  effective-theory approach, is expected to
be present. The inclusion of a bare contribution to the Planck mass in
the dual 4D theory is dual to including a UV-localized Einstein-Hilbert term in the 5D theory. In this case the 4D Planck mass is:
\bea
M_{Pl}^2\sim \frac{M_*^3}{k} + M_{UV}^2 =  \frac{M_*^3}{k} (1+v_0).
\eea
In the LWS limit one has $M_*\ll M_{Pl}$ and $v_0\gg1$ is necessary to
include 4D Einstein gravity, giving
$M_{Pl}^2\sim (M_*^3/k) v_0$. In the dual formulation it is clear why
it is consistent to include a large UV Einstein-Hilbert term,
$M_{UV}\sim M_{Pl}$, despite the fact that the UV cutoff is much less
than the Planck scale --- this is simply the standard low-energy
effective theory approach for including gravity. The theory breaks
down at energies $M_*\ll M_{Pl}$ and requires UV completion, but also
contains a consistent description of gravity for distances $>1/M_*$. 

It is also clear why the KK graviton spectrum is relatively
insensitive to the UV localized term. Including the UV term is equivalent to modifying the
fundamental sector of the dual theory. The spectrum of spin-two
composites of the CFT should not be significantly perturbed when
the properties of fields that are external to the CFT
(fundamentals) are altered. Similarly, the strength with which
the composite
dilaton interacts with other CFT modes should not depend on the
details of the fundamental sector, which is consistent with the
fact that the radion-IR-brane coupling
is insensitive to the large UV term. 

Sending the UV term to infinity, $v_0\rightarrow\infty$, makes the
effective 4D Planck mass infinite and decouples the massless
graviton from the particle spectrum. This
modification of the fundamental sector should not remove composite
modes from the spectrum, consistent with the finding that the radion
and the KK gravitons remain in the spectrum in the limit
$v_0\rightarrow\infty$. In the dual theory, interactions between
fundamental fields and the CFT are mediated by gravity. Thus, in the
limit that the massless graviton is removed, the two sectors decouple, in agreement
with our observation that the coupling of KK gravitons and the radion
to UV fields goes to zero for $v_0\rightarrow\infty$.
\section{Conclusion\label{sec:conc}}
We have studied the consistent inclusion of low-energy Einstein
gravity in theories constructed on a Little Warped Space; i.e.\ a
truncated slice of $AdS_5$ with a bulk gravity scale that is much less than
the 4D Planck mass, $M_*\ll M_{Pl}$. To provide
completely realistic low-energy theories,  models constructed on such spaces should 
reproduce 4D Einstein gravity for energies $E<M_*$. The approach we
have detailed provides a consistent description of low-energy gravity
by including a large UV-localized Einstein-Hilbert term (equivalently, a DGP term). The presence
of such a term can be motivated by string-theoretic realizations of
the RS model, in which a parametrically large UV curvature term
can arise~\cite{Brummer:2005sh}.  In
addition to realizing 4D Einstein gravity, we have detailed the gravitational sector of
the Little Warped Space, including both the KK graviton spectrum and
the properties of the radion. Furthermore, we have
shown that Goldberger-Wise stabilization can be
successfully implemented in these spaces, and obtained the dependence
of the radion mass and couplings on the brane localized
curvature.  These ingredients will play a role in
any theory  constructed on a LWS.\footnote{Models constructed on a LWS will require stabilization, though other possibilities exist beyond the Goldberger-Wise approach. However, the stabilization of the extra dimension in string realizations of RS models is well modeled by the Goldberger-Wise method~\cite{Brummer:2005sh}, providing further motivation for this approach.}

Our results demonstrate that realistic low energy theories can be constructed on Little Warped Spaces and, in particular, have relevance for
existing models in the literature~\cite{Davoudiasl:2008hx,McDonald:2010jm,Duerr:2011ks}. Given that large UV-localized curvature terms arise in string realizations of the RS model~\cite{Brummer:2005sh}, we believe
 that the study of phenomenological models on Little Warped
 Spaces is well
motivated --- both within the context of solutions to the SM
gauge-hierarchy problem, as in the Little RS
model~\cite{Davoudiasl:2008hx}, and the study of warped hidden sectors,
as in the Mini-Seesaw~\cite{McDonald:2010jm,Duerr:2011ks}.
\section*{Acknowledgements\label{sec:ackn}}
DG was supported by the Netherlands
Foundation for Fundamental Research of Matter (FOM), and the
Netherlands Organisation for Scientific Research (NWO), and thanks the  MPIK (Heidelberg) for hospitality. KM thanks H.~Davoudiasl and is grateful to Nikhef for hospitality.

\section*{Appendix}
\appendix
\section{Gauge Freedom with the Variables $P_i$\label{app:gauge}}
The action of a general coordinate transformation
$x^M\to x^M+\xi^M(x,y)$ on an arbitrary metric perturbation $h_{MN}$ is
\begin{equation}
\begin{aligned}
h_{\mu\nu} &\to h_{\mu\nu} - \nabla_\mu \xi_\nu - \nabla_\nu \xi_\mu - g_{\mu\nu}\frac{2a'}{a}\xi^5 \:,\\
h_{\mu5} &\to h_{\mu5} - g_{\mu\nu} {\xi^\nu}' - \partial_\mu \xi^5 \:,\\
h_{55} &\to h_{55} - 2 {\xi^5}' \:.
\end{aligned}
\label{h_gen_coord_trans}
\end{equation}
One can always perform a gauge transformation to bring the perturbation to a
straight gauge with the form of Eq.~(\ref{p_metric_ansatz}). Note that a straight gauge is defined by the demand that ($i$) the $(M,N)=(\mu,5)$ components of the metric vanish: $G_{\mu 5}=G^0_{\mu 5}+h_{\mu 5}=0$, where $G^0_{MN}$ is the background metric; and that ($ii$) the gauge parameters $\xi^5(x,y)$ vanish at the boundaries, $\xi^5(x,y_i)=0$ for $y_i=0,L$~\cite{Carena:2005gq}. The latter restriction ensures that the boundaries are located at the slices $y_i=0,L$, and removes the pure gauge brane-bending mode from the spectrum~\cite{Carena:2005gq}. 

Starting with an arbitrary $h_{55}$, one obtains the form used in (\ref{p_metric_ansatz}) by performing a gauge transformation with parameters $\xi^\mu(x,y)=0$ and:
\bea
\xi^5(x,y)=\frac{1}{2}\int_0^yd\tilde{y}\ (h_{55}-2P_1) + \frac{1}{2}\left[a^2P_3'\right]^y_0,
\eea
subject to the constraint:
\bea
\int_0^Ld\tilde{y}\ P_1- \frac{1}{2}\left[a^2P_3'\right]^L_0 = \frac{1}{2}\int_0^Ld\tilde{y}\ h_{55},
\eea
in accordance  with the definition of a straight gauge. Note that, if the boundary conditions require $P_3'\propto P_1$, as found in the text [see Eq.~(\ref{p1p3_boundary_condition})], we can write $P_1=p_1(y)\psi(x)$ and obtain
\bea
\psi(x)=\frac{\frac{1}{2}\int_0^Ld\tilde{y}\ h_{55}}{\int_0^Ld\tilde{y}\ p_1-\frac{1}{2} \left[a^2b_i v_ip_1\right]^L_0},
\eea 
as the definition for $\psi$, where $b_i$ are constants. Thus, for arbitrary $h_{55}$, the denominator should be nonzero. One completes the transition to a straight gauge by performing  a second gauge transformation with $\xi^5(x,y)=0$ and $\xi^\mu(x,y)$ such that:
\bea 
h_{\mu 5}-2g_{\mu\nu}{\xi^{\nu}}'=0.
\eea
The form of the $(M,N)=(\mu,\nu)$ components of the perturbation can then always be cast as in Eq.~(\ref{p_metric_ansatz}).

Further gauge transformations will preserve the
straight-gauge form provided the gauge parameters $\xi^M$ satisfy 
\begin{equation}
\partial_\mu\xi^5(x,y) = -g_{\mu\nu} \partial_5 \xi^\nu(x,y) \:,
\end{equation}
which gives
\begin{equation}
\xi^\mu(x,y) = -\int^y d\tilde{y}\  g^{\mu\nu}\partial_\nu \xi^5(x,\tilde{y})  + \chi^\mu(x) \:.
\end{equation}
Thus, in a general straight gauge one has a remaining gauge choice described by
the arbitrary functions $\xi^5(x,y)$ and $\chi^\mu(x)$, with
$\xi^\mu(x,y)$ fixed as above and $\xi^5|=0$.  Note that,
because we have not chosen $h_{55}$ to be of separable form, we
presently have more remnant gauge freedom than~\cite{Carena:2005gq}
(see Eq.~(226) in the published version). Specifically, once $h_{55}$ is chosen separable, the $x$-dependence of $\xi^5(x,y)$ is fixed.

In a straight gauge, the transformation \eqref{h_gen_coord_trans} becomes
\begin{align}
h_{\mu\nu} &\to h_{\mu\nu}
  - \nabla_\mu \chi_\nu(x) - \nabla_\nu \chi_\mu(x)
  - g_{\mu\nu}\frac{2a'}{a}\xi^5(x,y)
  + 2 a^2 \nabla_\mu\nabla_\nu \int^y d\tilde{y} \, \frac{\xi^5(x,\tilde{y})}{a^2}
  \:,\nonumber\\
h_{\mu5} &\to h_{\mu5} \:,\\
h_{55} &\to h_{55} - 2 \partial_5 \xi^5(x,y) \:.\nonumber
\end{align}
Consider the action of a gauge transformation on the variables
$P_i$ employed in the text.
Let us ignore the 4D $\chi^\mu(x)$ general coordinate transformation
and concentrate on the $\xi^5(x,y)$ gauge freedom.
If we choose
\begin{align}
P_1 &\to P_1 \:,\nonumber\\
P_2 &\to P_2 \:,\label{app:aq:pi_trans}\\
P_3 &\to P_3 + 2\int^y d\tilde{y} \, \frac{\xi^5(x,\tilde{y})}{a^2} \:,\nonumber
\end{align}
then from~\eqref{p_metric_ansatz} we have
\begin{align}
h_{\mu\nu} &= a^2\nabla_\mu\nabla_\nu P_3 + a^2\eta_{\mu\nu}\left(2P_2-aa'P_3'\right) \nonumber\\
&\to h_{\mu\nu} - g_{\mu\nu} \frac{2a'}{a} \xi^5(x,y) + 2a^2\nabla_\mu\nabla_\nu \int^y d\tilde{y} \, \frac{\xi^5(x,\tilde{y})}{a^2}
\end{align}
and
\begin{align}
h_{55} &= 2P_1-(a^2P_3')' \nonumber\\
&\to 2P_1-\partial_5\left[a^2\left(P_3'+\frac{2\xi^5(x,y)}{a^2}\right)\right] \\
&= h_{55} - 2\partial_5\xi^5(x,y) \nonumber\:.
\end{align}
This is the correct transformation of the metric perturbations, which
tells us that the transformation (\ref{app:aq:pi_trans}) correctly
encapsulates the remaining gauge freedom.
Thus, $P_1$ and $P_2$ remain unchanged under a gauge transformation with
non-zero $\xi^5(x,y)$, while $P_3$ changes as above. 

It would seem that one can choose $\xi^5(x,y)$ such that $P_3$ is gauged from the spectrum. However, this is not necessarily the case. Noting that under a gauge transformation one has:
\bea
P_3'\rightarrow P_3'+\frac{2\xi^5(x,y)}{a^2},
\eea
and that straight gauges demand $\xi^5|=0$, we see that $P_3'$ can only be gauged away when $P_3'|=0$. For $b_i\ne 0$, as found in the text, this is only true for $v_i=0$. 

One can use the remaining gauge freedom to relate $P_3'$ to $P_1$ by specifying the $x$-dependence of $\xi^5$ such that
\bea
P_3'(x,y)\rightarrow F_3(y) P_1(x,y),
\eea
with $F_3(y)$ satisfying $F_3|=b_iv_i$ but otherwise arbitrary. The remaining gauge freedom, beyond the 4D general coordinate transformations, is then fixed by specifying the $y$-dependence of $\xi^5$ to determine $F_3$.



\end{document}